# The CTOF measurements and Monte Carlo analyses of neutron spectra for backward direction from iron target irradiated by protons with energies from 400 to 1200 MeV

I.L. Azhgirey, V.I. Belyakov-Bodin, I.I. Degtyarev

*Institute for High Energy Physics, Protvino, Russian Federation*

S.G. Mashnik

*Los Alamos National Laboratory, Los Alamos, NM 87545, USA*

F. X. Gallmeier, W. Lu

*Oak Ridge National Laboratory, Oak Ridge, TN 37830, USA*

## Abstract

A calorimetric-time-of-flight (CTOF) technique was used for real-time, high-precision measurement of neutron spectrum at the angle of $175^o$ from the initial proton beam direction, which hits a face plane of a cylindrical iron target of 20 cm in diameter and 25 cm thick. A comparison was performed between the neutron spectra predicted by the MARS and the MCNPX code systems and measured by experiments for 0.4-, 0.6-, 0.8-, 1.0-, and 1.2-GeV protons.

## Introduction

Detailed description of the experimental facility construction, the calorimetric-time-of-flight (CTOF) technique, method of data taking, and experimental results for tungsten target irradiated by protons have recently been described in Ref. [1,2]. Therefore, only a brief description of the ZOMBE facility and the CTOF technique needed for understanding the problem is presented in this paper.

The MARS and the MCNPX code systems are useful tools for the analysis and design of systems incorporating high-intensity neutron sources such as accelerator-driven system (ADS). These codes simulate the transport of particles in matter with the cascading of secondary particles over the wide energy range. Encompassed in this is the ability to determine the neutron spectrum around a system. For many system designs, and especially for high-energy, high beam-current applications, an important design factor is the neutron spectrum emitted from the ADS target. Consequently, it is important that the accuracy for neutron spectrum predictions by the codes be well determined. The goal of this study is to determine the precision of the MARS and the MCNPX codes in making these predictions at a wide angle.

The geometry for both codes that modeled the experimental configuration consists of a cylindrical target and a detecting volume.

## Facility description and operation principle

We used a proton beam, which was extracted from the U-1.5 ring accelerator (a booster) in the Institute of High Energy Physics, Protvino, Russian Federation by a tripping magnet. The booster cycle time was about 9 s containing 29 pulses of about 30 ns long, each containing $1.3 \times 10^{11}$ protons. Extraction time is 1.6 s. Beam intensity measurements were carried out for each pulse by using an induction current sensor [3] with an accuracy of about 7%. The proton beam energy values were determined with an error of approximately 0.5% in the proton energy value. The average radial distribution of protons in the beam has been obtained [1,2] for this experiment and was compatible with a Gaussian of a full-width at half-maximum of 24 mm. A proton beam hits a face plane of target placed on the ZOMBE facility (see Fig. 1). The target axis position was superimposed on the geodesic beam axis.

The CTOF technique consists of a set of current mode time-of-flight instruments each having a set of organic scintillation detector placed at a different distance from a source at an interested angle from the initial proton beam direction. One - litre (10 x 10 x 10 cm$^3$) organic scintillators were used as the detectors. The first detector was placed at distance 8.7 m from the face plane of the target at 175$^o$ to the proton beam direction. The second detector was placed at distance 29 m at the same angle. Wave-length shifters were coupled with the scintillators and its output signals were transported to each photo-multiplier tube (PMT) by a 120-m-long light-guide (LG). The outputs of PMT were digitized by a TDS 3034b oscilloscope (0.4 ns resolution and 200 MHz bandwidth). The data were obtained within a time of 2000 ns from the irradiation cycle start and were sent to an off-line PC computer for further analyses. All of the beam pickup signals were digitized so for reconstruction of average proton beam intensity as a function of time. In measurement, 5000 proton beam pulses were used to average photometric signals.

To exclude any radiation-induced noise from photometric signal of detectors, such as neutrons reflected from walls, ceiling and floor, radiations from accelerator, and so on, it was necessary to make additional measurement for the case when the burst source overshadowed for detectors by cylindrical shield. We used the shield from a set of 45-cm-thick iron and 40-cm-thick tungsten discs, both 20 cm in diameter.

Difference of these two measurements will give a pure signal from the source.

Amplitude of a light signal in a detector caused by a nonzero-mass particle is:

$$u\left(t\right) = k \cdot S\left(E\right) \cdot e_f\left(E\right) \cdot dE / dt \,,$$

where $S(E)$ is the energy spectrum of particles emitted from a target, $e_f(E)$ is the sensitivity of the detector, $dE/dt$ is the time derivative of particle energy for $r$-distance detector, and $k$ is a coefficient. The sensitivity $e_f(E)$ of detector to neutron (and cosmic μ-meson ($e\mu$) was calculated by the RTS&T code [4] (see Fig. 2). We used time integral of a photometric signal of scintillator caused by a cosmic μ-meson signal for determination of $k$-coefficient, i.e. for absolute calibration of detectors.

Digitized voltage amplitude is determined by the following integral equation:

$$U\left(t\right) = u\left(t\right)^* \cdot p\left(t\right)^* \cdot d\left(t\right)^* \cdot LG\left(t\right)^* \cdot PMT\left(t\right) \quad , \tag{1}$$

where $p(t)$ is proton beam intensity as a function of time; $d(t)$, $LG(t)$, and $PMT(t)$ are pulse functions of detector, LG, and PMT, respectively; and the asterisk (*) is convolution transform symbol:

$$u\left(t\right)^* \cdot p\left(t\right) = \int_o^t u(\xi) \cdot p\left(t - \xi\right) d\xi \,.$$

An original code was made for determining $u(t)$ function from the Eq. (1) by using $U(t), p(t), d(t), LG(t),$ and $PMT(t)$.

The part of photometric signal caused by gamma rays from the target was extracted by the following manner. The time structure of this part of signal does not depend on the distance between the source and detector for experimental condition. Therefore, we determined this shape from the distant detector signal, which could be separated easily and clearly into gamma rays and nucleon components for the detector with a lead filter. The part of photometric signal caused by proton flux can be extracted from the total photometric signal of a detector by inserting different kinds of filters between the detector and the spallation source. In the experiment, we used two detectors at 29-m distance, each with and without a lead filter. The photometric signal from the detector without a filter was subtracted of the gamma- and proton-induced signals to obtain neutron component up to 13 MeV. It is possible to reduce the relative error for low-energy part of these signals and expand energy range, for example, by improving the light system and by using a higher-digitizing oscilloscope. Nevertheless, measurements were conducted also by using detector at 8.7-m to get neutron spectrum below 13 MeV, because it was easier.

## Monte Carlo simulation

The MCNPX calculated backscattering neutron spectra as shown in Figures 3-8 were performed with the 2.6.0 version of MCNPX [5] using the Bertini INC [6] coupled with the Dresner evaporation model [7] (MCNPX defaults, using also the Multistage Preequilibrium Model (MPM) [8]), the Liege INC model INCL4 [9] merged with the GSI evaporation/fission model ABLA [10], as well as the 03.01 version of the Cascade-Exciton Model event-generator CEM03.01 [11, 12] using an improvement of the Generalized Evaporation Model GEM2 by Furihata [13] to calculate evaporation/fission and its own Modified Exciton Model (MEM). In addition, a newer version of MCNPX, MCNPX 2.7.A [14] was applied using the latest implementation of CEM, CEM03.02 [12, 15]. Let us note in advance that all neutron spectra calculated here with MCNP 2.6.0 using CEM03.01 are very similar to the ones obtained with MCNPX 2.7.A using CEM03.02, therefore we present below only results by CEM03.02. For the proton and neutron transport and interactions up to 150 MeV, MCNPX in all cases applied tabulated continuous-energy cross sections from the LA150 proton and neutron libraries. The calculation assumed a Gaussian-distributed incident proton beam with a FWHM of 2.4 cm hitting a cylindrical iron target. The radius of the iron target is 10 cm and the length is 25 cm. The backscattered neutrons were tallied at an angle of $175^{\circ}$ of the incident beam direction and at 6 m upstream of the front surface of the target. A ring surface tally was adopted to improve the efficiency of the simulation taking advantage of the symmetry of the problem.

The MARS code [16] simulates a process of development on the nuclear-electromagnetic cascades in matter. Its physical module is based mostly on parameterization of the physical processes description. It provides flexibility and a rather high operation speed of the code for engineering applications and optimization tasks. MARS is in use for radiation-related modeling at accelerators, such as shielding design [17], dose distribution and energy deposition simulations [18], and radiation background calculations [19]. In this region of the primary proton energy (below 5 GeV), MARS uses a phenomenological model for the production of secondary particles in inelastic hadron-nucleus interactions [20]. For low-energy neutron transport at energies below 14.5 MeV threshold and down to the thermal energy, MARS uses a multi-group approximation and a 28-group library of neutron constants.

The energy threshold for the charged hadron transport is 10 MeV. The target was considered as solid iron cylinder with a length of 25 cm and diameter of 20 cm. A detecting volume (10x10x10 cm$^3$) was placed in 9 m upstream of the target front face plane, and the angle between beam direction and direction from target to detector is 175°. A Gaussian distribution with $\sigma_{x,y}$ = 1.4 cm of proton beam  was used, which is an approximation of the measured beam distribution. Test runs show no difference in spectrum between point-like beam and realistic distribution. Space between target and detector was filled with air.

## Comparison of experimental and calculation results

As shown in Figs. 3-7, MARS results agree well with the measured spectra and the MCNPX results by all event-generators used here also are in a reasonable agreement with all measured spectra. MARS overestimates a little the 20-80 MeV portion of spectra at 400 MeV and underestimates the high-energy tails of spectra at 800, 1000, and especially at 1200 MeV. All event-generators of MCNPX underestimates a little the measured spectra in the ~ 5-10 MeV energy region, as well as at high energies, at the very end of the spectra tails, and also a little, at very low energies, in the 1-3 MeV region. The overall agreement of results by different event-generators with the measured spectra depends mostly on the proton incident energy: CEM tends to agree better with the data at high energies of 1 and 1.2 GeV, while Bertini+Dresner and INCL+ABLA agree better with the data at lower incident energies.

A little difference is observed for neutron energies from several MeV to ~50 MeV, where a shoulder is presented on the measured spectra around 5 MeV showing higher backscattered neuron flux compared to the MCNPX calculated spectra for neutrons from several MeV to 10 MeV and lower neutron flux for neutrons above 10 MeV up to ~50 MeV. Neither event-generators (CEM03 and BERTINI+Dresner) using there own, but different preequilibrium models nor the INCL+ABLA, which does not use a preequilibrium stage, are able to capture this shoulder. Such a discontinuity (as represented by the shoulder) on the measured spectrum probably indicates the intersection of the flux from the low-energy evaporation neutrons and high-energy cascade neutrons. Although the high-energy cascade neutron flux calculated by all models of MCNPX agrees reasonably well with the measurement at incident proton energies above 400 MeV, all the MCNPX event-generators used here underestimate slightly the measured spectrum at 400 MeV, up to a factor of two.

MARS underestimates the high-energy tails of spectra at highest beam energies. It produces approximately 3 times less number of neutrons compare to the data. Such behavior for this kinematical region is similar to the tungsten target case.

## Acknowledgements

We are grateful to G.A. Losev and A.V. Feofilov (both of the Institute for High Energy Physics) for providing the beam so reliably under the conditions demanded for this experiment. This work was partially supported by the US DOE.

# List of Figures

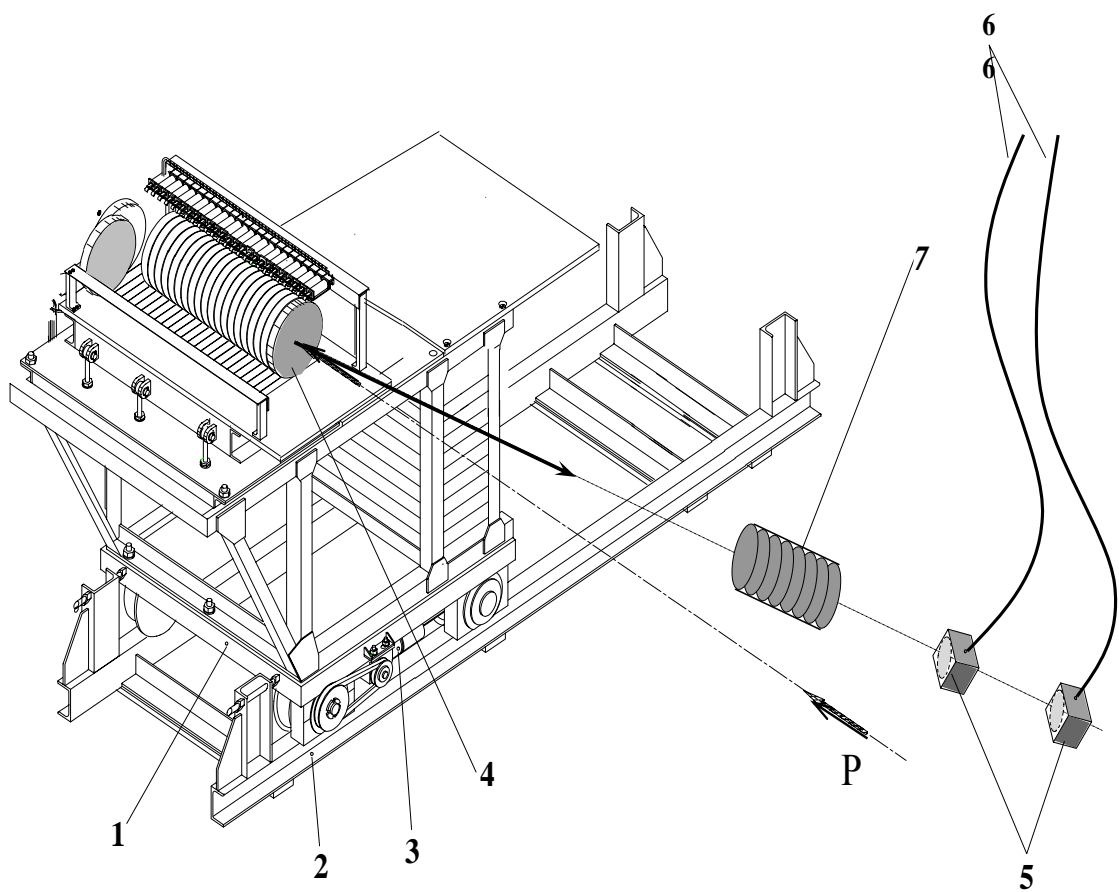

P

Fig. 1

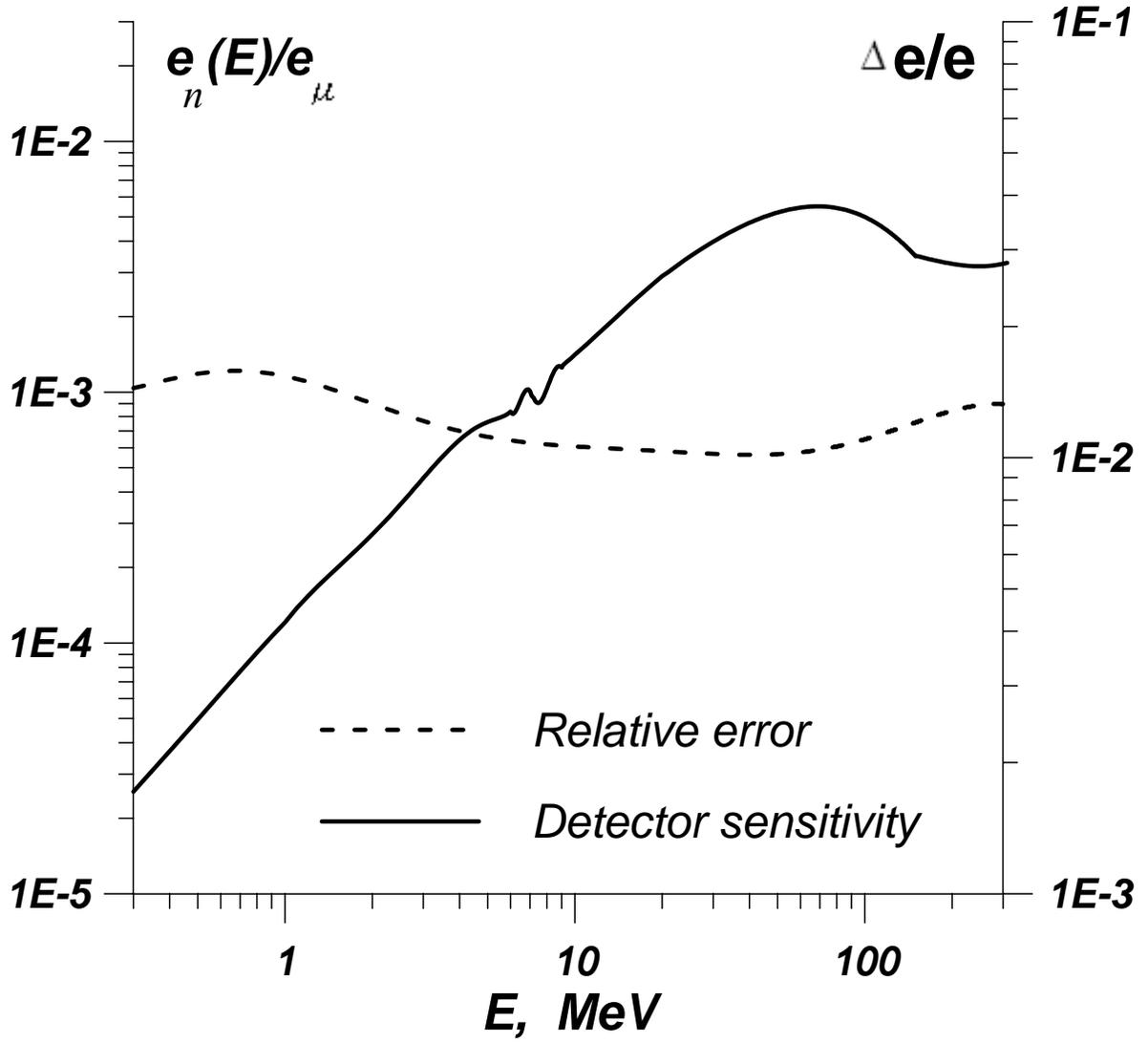

Fig. 2

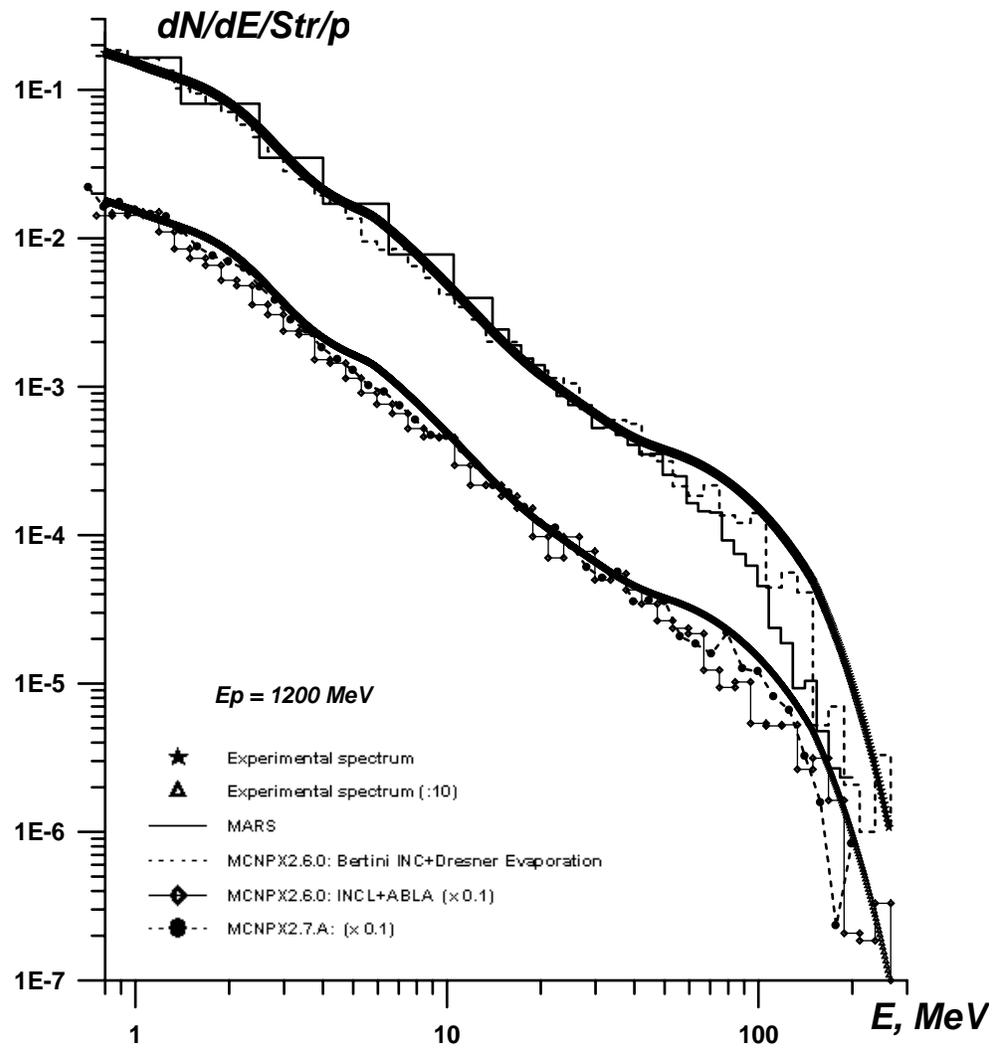

Fig. 3

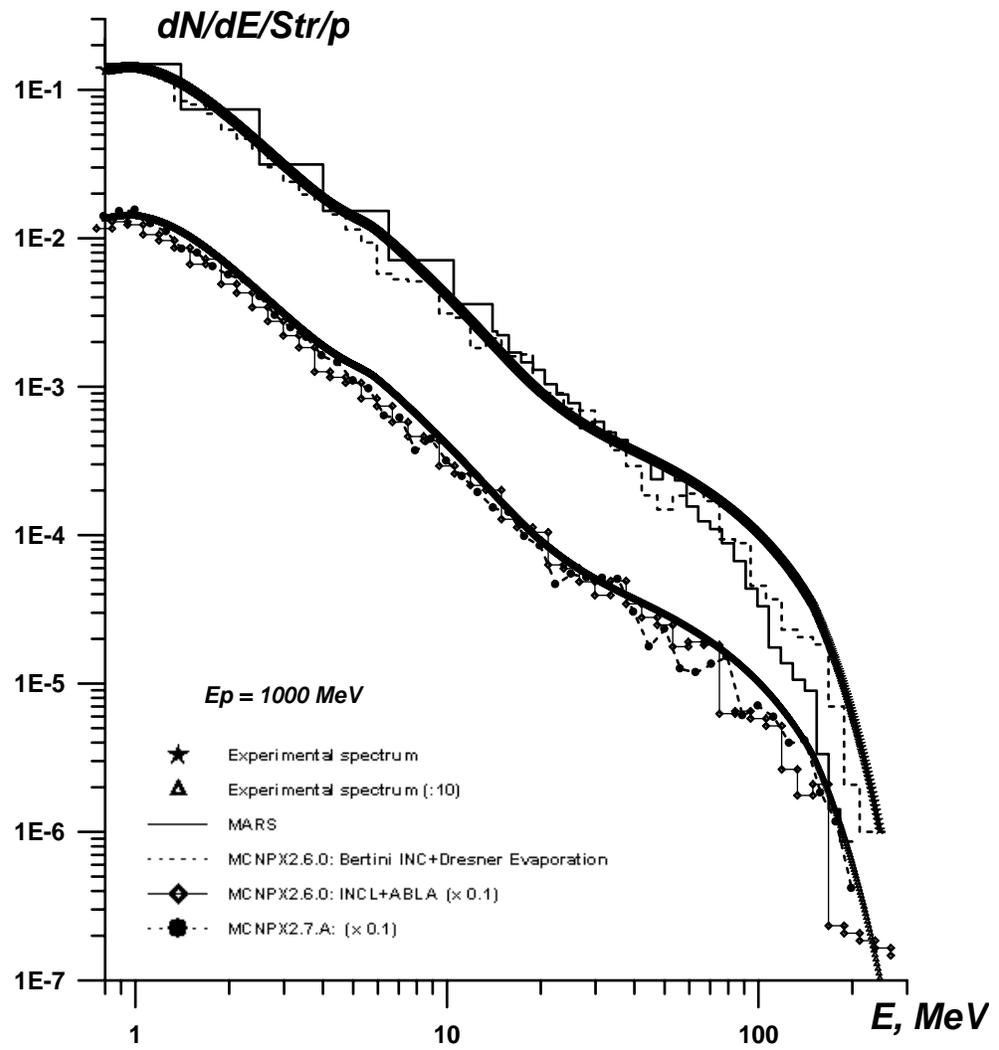

Fig. 4

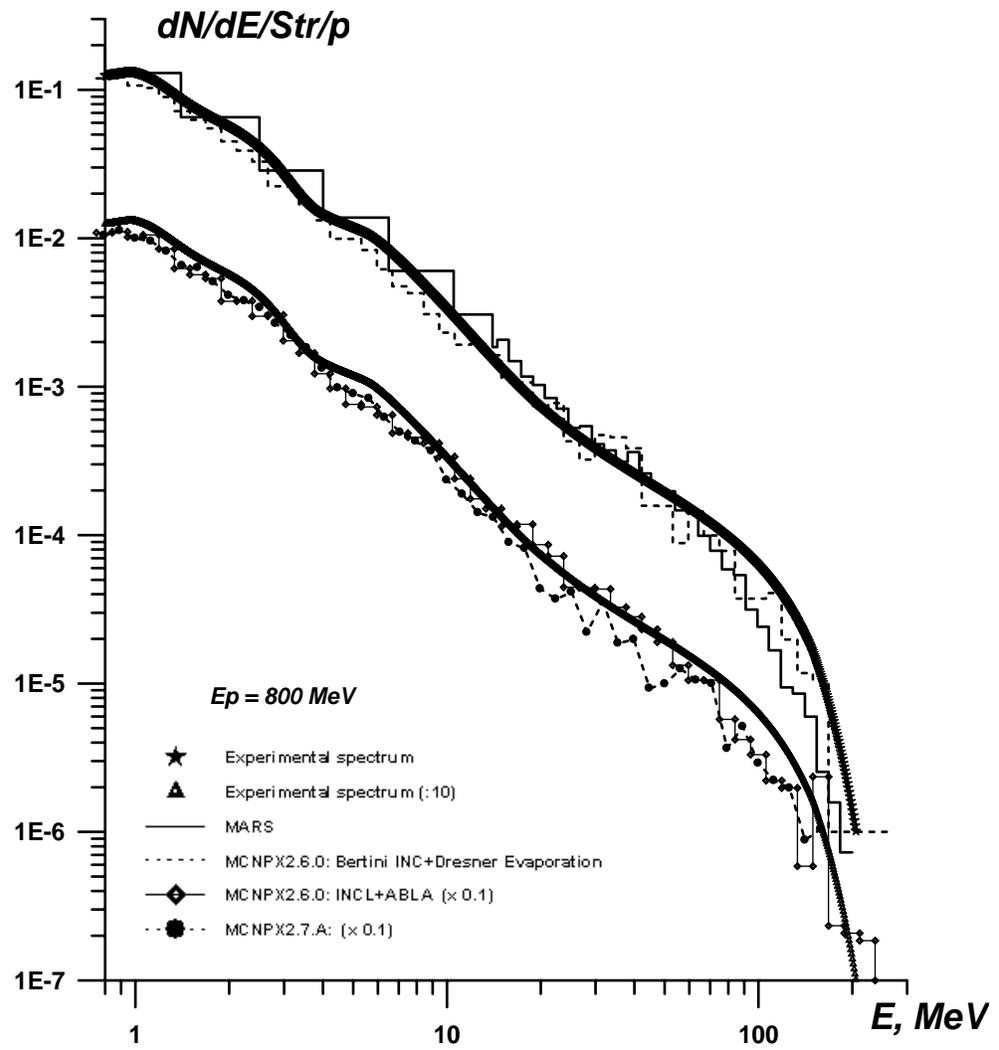

Fig. 5

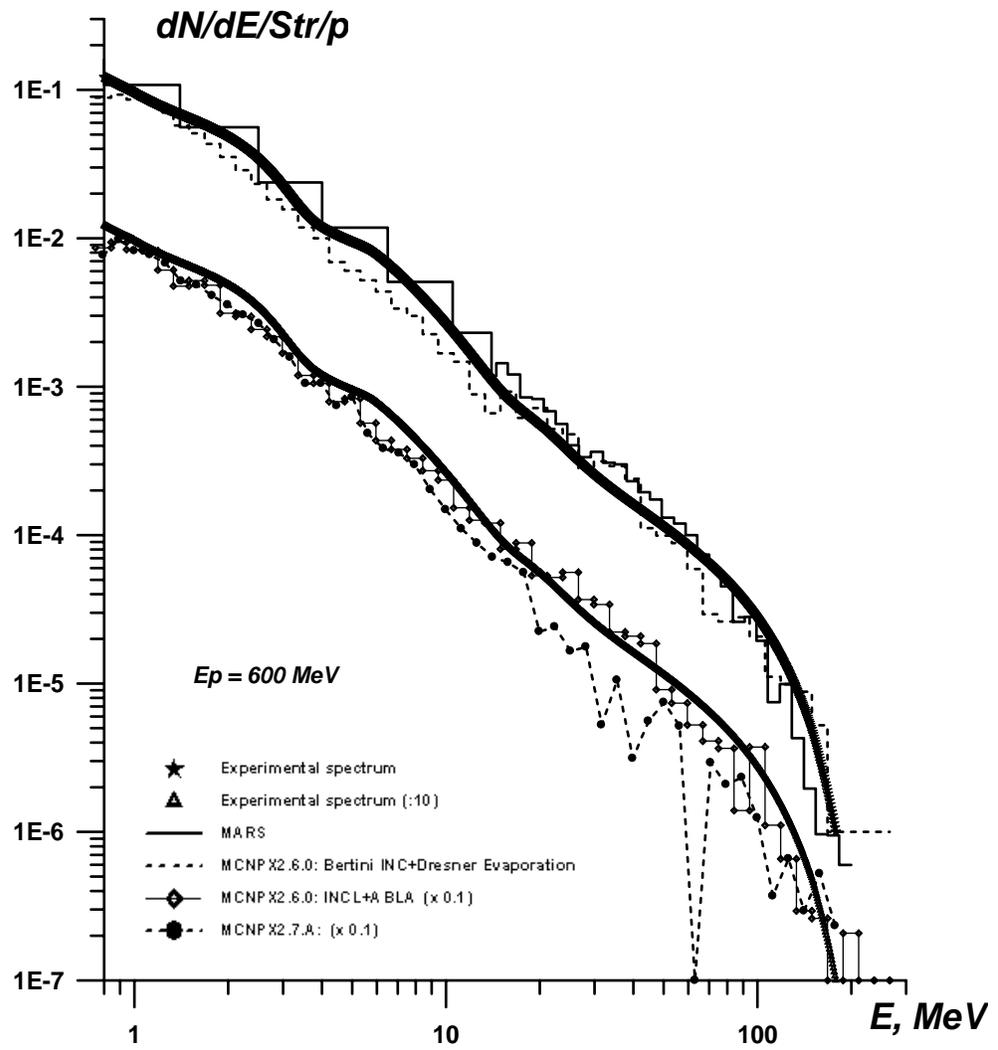

Fig. 6

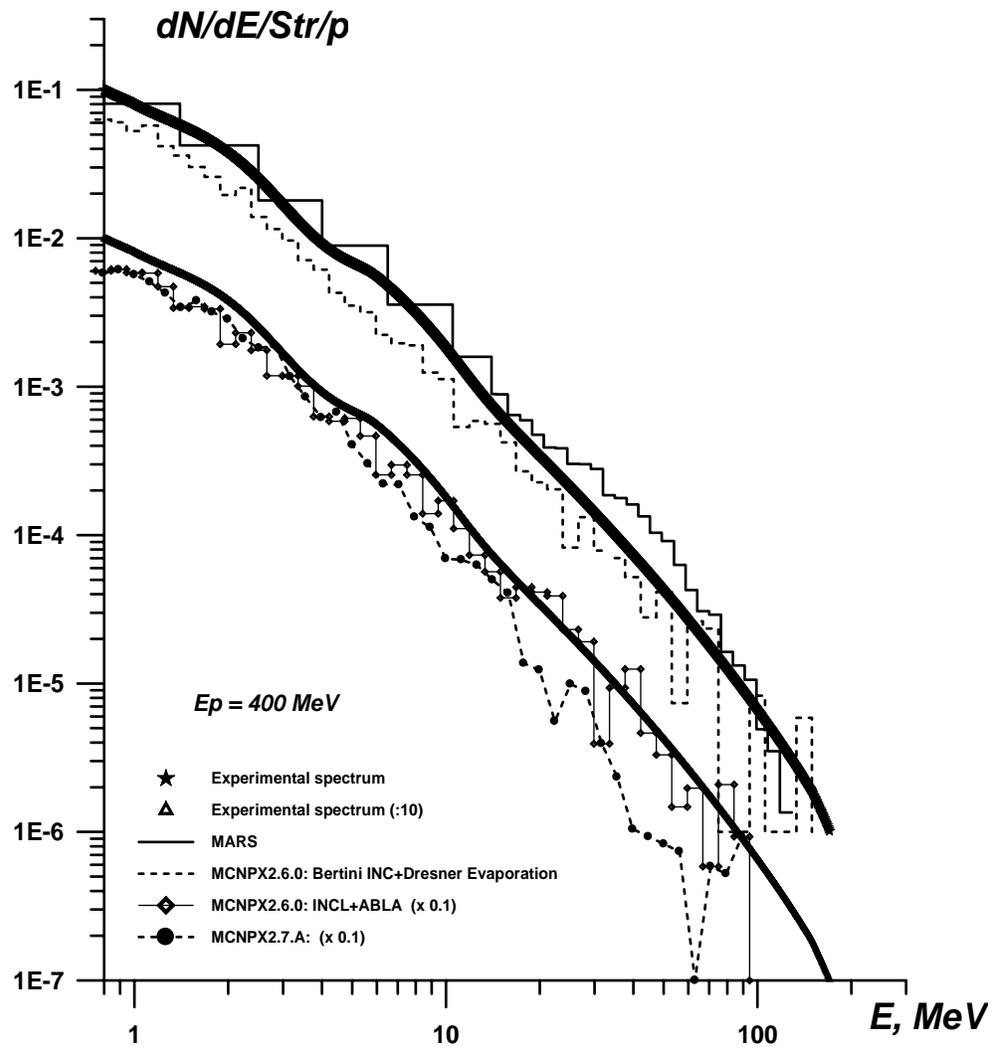

Fig. 7